\documentclass[%
 aip,
 amsmath,amssymb,
 reprint,%
]{revtex4-2}

\usepackage{graphicx}
\usepackage{dcolumn}
\usepackage{bm}

\usepackage[utf8]{inputenc}
\usepackage[T1]{fontenc}
\usepackage{mathptmx}
\usepackage{etoolbox}
\usepackage[usenames,dvipsnames]{color}
\usepackage{soul}
\usepackage{cancel}

\makeatletter
\def\@email#1#2{%
 \endgroup
 \patchcmd{\titleblock@produce}
  {\frontmatter@RRAPformat}
  {\frontmatter@RRAPformat{\produce@RRAP{*#1\href{mailto:#2}{#2}}}\frontmatter@RRAPformat}
  {}{}
}%
\makeatother
\begin{document}

\preprint{AIP/123-QED}

\title{Origin of electrical noise near charge neutrality in dual gated graphene device}
\author{Aaryan Mehra}
\author{Roshan Jesus Mathew}
\author{Chandan Kumar}%
 \email{kchandan@iisc.ac.in}
\affiliation{ 
Centre for Nano Science and Engineering, Indian Institute of Science, Bangalore-560012, India 
}%

\date{\today}

\begin{abstract}
This letter investigates low frequency $1/f$ noise in hBN encapsulated graphene device in a dual gated geometry. The noise study is performed as a function of top gate carrier density ($n_{TG}$) at different back gate densities ($n_{BG}$). The noise at low $n_{BG}$ is found to be independent of top gate carrier density. With increasing $n_{BG}$, noise value increases and a noise peak is observed near charge inhomogeneity of the device. Further increase in $n_{BG}$ leads to decrease in noise magnitude. The shape of the noise is found to be closely related to charge inhomogeneity region of the device. Moreover, the noise and conductivity data near charge neutrality shows clear evidence of noise emanating  from combination of charge number and mobility fluctuation.
\end{abstract}

\maketitle

Graphene, a single sheet of carbon has emerged as one of the most promising candidate for the future device applications which can supersede the silicon technology. However, graphene devices are very sensitive to disorder fluctuations which leads to fluctuation in channel current\cite{balandin2013low,islam2022benchmarking}. Thus, quantifying the effect of the disorder becomes crucial for using the graphene based devices for practical applications.  The low frequency $1/f$ noise technique has been used extensively to study the effect of various  disorders like roughness and trap states in the substrate\cite{kaverzin2012impurities}, short and long range disorder\cite{zhang2011mobility}, metal contact to graphene\cite{karnatak2016current}, charge inhomogeneity\cite{heller2010charge}, grain boundaries\cite{kochat2016magnitude} etc. The $1/f$ noise not only affects the low frequency $<100 kHz $ but its possible up-conversion also affects the phase noise at higher frequency.  Thus, understanding the noise spectrum in graphene devices becomes cardinal for technological applications.\\

However, even after a more than a decade of extensive research, there is no consensus on the origin and mechanism of $1/f$ noise in graphene devices\cite{balandin2013low,kayyalha2015observation,kumar2016tunability,kakkar2020optimal}. The noise study in graphene has been performed on both silicon\cite{cultrera2018role,wu2018low} and hBN substrate \cite{stolyarov2015suppression,kayyalha2015observation,kumar2016tunability,kakkar2020optimal}. The study reveals that the graphene devices on hBN substrate shows much smaller noise as compared to the graphene on a silicon substrate. The noise reduction in hBN encapsulated devices are associated to the  better screening of graphene from the trap states in silicon substrate and to very smooth surface of hBN which is free of dangling bonds\cite{stolyarov2015suppression,kumar2016tunability,kayyalha2015observation,kakkar2020optimal}. However, noise in hBN encapsulated graphene devices show different shapes and magnitude as a function of carrier density\cite{stolyarov2015suppression,kumar2016tunability,kayyalha2015observation,kakkar2020optimal}.  Moreover, the noise magnitude varies from device to device and also depends on the type of encapsulating hBN - commercial hBN or NIMS hBN\cite{kakkar2020optimal}. Furthermore, it is  reported that noise magnitude in hBN encapsulated graphene can be further reduced by adding an additional graphite gate\cite{kakkar2020optimal}.  Thus, there is an increasing evidence that carrier number fluctuation is the major noise source in graphene devices both on silicon and hBN encapsulated devices. However, there are also evidences of noise stemming from mobility fluctuation as well. The claim is made on the basis of electron-beam irradiated graphene\cite{zahid2013reduction,cultrera2018role} on silicon substrate and from non-monotonous variation of noise with magnetic field  in hBN encapsulated\cite{rehman2022nature} and suspended\cite{kamada2021suppression} graphene device.\\

In this letter, we present a schematic study of low frequency $1/f$ noise in hBN encapsulated graphene device in a dual gated geometry. We have performed noise study in a small region of graphene which is underneath the top gate. The noise measurements are performed as a function of $n_{TG}$ at different $n_{BG}$. We find that noise is small and almost independent of $n_{TG}$ at small values of $n_{BG}$. With increasing  $n_{BG}$, noise starts to increase and obtains a maximum value near $n_{BG} = 1.5 \times 10^{12} cm^{-2}$. With the further increase in $n_{BG}$, noise value decreases and becomes independent of $n_{BG}$. Moreover, our measurements shows that noise amplitude is minimum at the charge neutrality point (CNP) and it increases on both side with electron and hole doping  with its peak appearing close to the charge inhomogeneity of the device. Our results shows that noise in hBN encapsulated graphene can be tuned by nearly two orders of magnitude. Furthermore, our analysis of noise and channel conductivity  provides a clear evidence of  noise emanating from combination of charge number and mobility fluctuation. \\
\begin{center}
	\begin{figure*}[t]
		\includegraphics[width=1.0\textwidth]{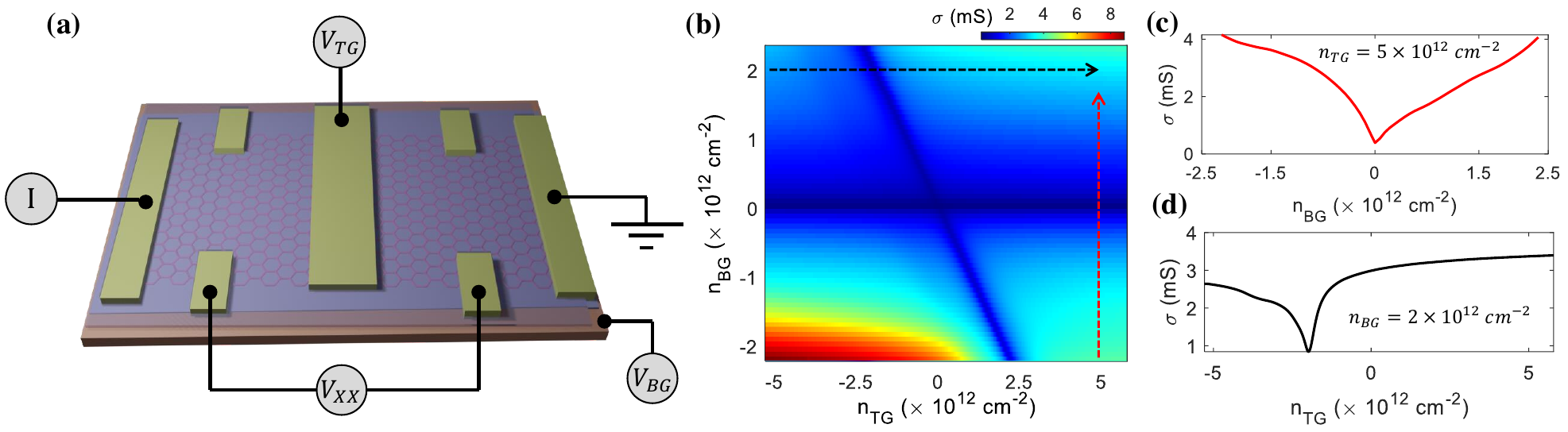}
		\caption{(color online) (a) Schematic of the measured hBN encapsulated graphene device along with resistance measurement scheme. The device is fabricated using dry transfer technique\cite{zomer2011transfer}. The graphene is encapsulated between the top and bottom hBN with thickness $\sim $14 nm and $\sim$25 nm, respectively. The back gate controls the carrier density throughout the device while the top gate controls the carrier density only beneath the top gate region. (b) 2D color map of conductivity as a function of top gate and back gate carrier density. The horizontal blue stripe shows the Dirac point throughout the device while the diagonal stripe shows the Dirac point from a small section of graphene underneath the top gate region. The vertical and horizontal cut line from panel (b) is presented in (c) and (d), respectively. As can be seen, the conductivity is minimum at the Dirac point and increases with electron or hole doping. All the measurements are performed at 77 K.}
		\label{fig:images}
	\end{figure*}
\end{center}

The hBN encapsulated graphene device is fabricated by following standard dry transfer technique\cite{zomer2011transfer}. In brief, a glass slide is prepared with Poly methyl methacrylate (PMMA) layer and graphene is exfoliated on it. hBN is exfoliated on a freshly cleaved highly p doped silicon wafer. The glass slide containing PMMA and graphene is loaded in a micro-manipulator and graphene is transferred on hBN. This is followed by cleaning the stack of hBN/graphene in acetone followed by IPA cleaning. The standard electron beam lithography is used to pattern contacts on the graphene heterostructure, followed by thermal evaporation of 5 nm chromium  and 70 nm gold at the base pressure of 3e-7 mbar. The device is annealed in vacuum for 3 hours at 400 $^\circ$ C. To define the top gate,
a thin top hBN is transferred on the hBN/graphene stack. This is followed by final step where electron beam lithography is used to pattern the contacts, followed by metal deposition.\\

Figure 1 (a) presents the schematic of our device along with resistance measurement scheme. The optical image of the device is shown in Supplementary material (SM) Fig. S1. All the measurements are performed at 77 K.  The graphene is encapsulated between the top and bottom hBN. The carrier density in the device is tuned by the combination of top gate and back gate voltages. The thin top hBN and 300 nm thick $SiO_2$ act as the top and back gate dielectric, respectively. The back gate controls the density throughout the graphene channel while the top gate tunes the carrier density in a small portion of graphene, underneath the top gate. The total channel length and width is $8 \mu m$ and $2 \mu m $, respectively. The distance between voltage probes are $ 4 \mu m$ and the top gate width is $2 \mu m $. Figure 1 (b) shows the four probe conductivity data as a function of top gate and back gate carrier density. A horizontal and diagonal stripe can be seen in the figure where the conductivity is minimum.  The horizontal stripe corresponds to Dirac point throughout the graphene channel which is independent of top gate carrier density while the diagonal stripe represents the Dirac cone in a small portion of graphene under the top gate region. Doping of carrier density (either electron or hole) leads to an increase in conductivity. This is shown explicitly in figure (c) and (d) by taking horizontal and vertical cut lines from figure 1(b), denoted by dashed arrow.  We estimate the top hBN thickness to be $\sim 14 nm$.\\

\begin{center}
	\begin{figure*}[t]
		\includegraphics[width=1\textwidth]{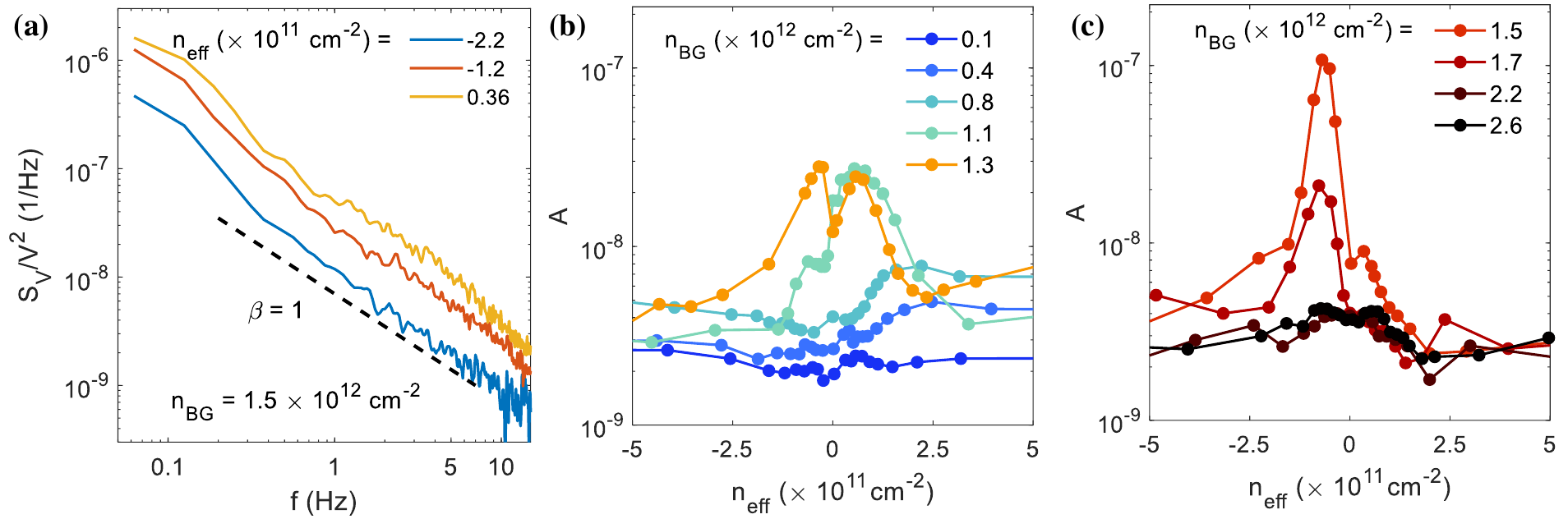}
		\caption{(color online) (a) Normalized noise spectral density in hBN encapsulated graphene device at few representative $n_{eff}$ at $n_{BG} = 1.5 \times 10^{12} cm^{-2}$. The dashed line represents a pure $1/f$ line with $\beta = 1 $. (b-c) Noise amplitude (A) as a function of effective carrier density $n_{eff} = n_{TG} - n_{Dirac}$ underneath the top gate at different $n_{BG}$. At low $n_{BG}$ the noise amplitude is almost independent of $n_{eff}$. With increasing $n_{BG}$, noise increases and maximum noise is observed at $n_{BG} = 1.5 \times 10^{12} cm^{-2}$. The further increase in $n_{BG}$, leads to reduction in noise amplitude.}
		\label{fig:images}
	\end{figure*}
\end{center}

The low frequency noise is a versatile tool to study the effect of metal contact, charge fluctuation, screening etc. on the device conductivity which can not be accessed by the traditional conductivity measurement methods\cite{balandin2013low,islam2022benchmarking}. It is characterized by normalized power spectral density which is inversely proportional to the frequency, $S_V/V^2 \varpropto f^{-\beta}$, with $\beta = 1$. In most of the semiconducting devices the noise power spectral density follows an empirical relation\cite{balandin2013low,islam2022benchmarking}:\\
\begin{equation}
S_V/V^2 = \alpha_H / Nf^{\beta}.
\end{equation} 
where $S_V/V^2$ is the normalized voltage noise density, $f$ is the frequency, $V$ is the bias voltage, $N$ is the total number of charge carriers and $\alpha_H $ is the Hooge's parameter.\\

Figure 2 (a) shows the normalized voltage noise density ($S_V/V^2$) at fixed $n_{BG}$ for a few representative $n_{eff}$. The measurement setup is discussed in SM.  A clear $1/f$ noise behaviour with $\beta \thickapprox  1$ can be seen from the figure. The noise value is quantified by calculating the noise amplitude (A) which is defined as\cite{balandin2013low}:

\begin{equation}
A = \frac{1}{N} \sum_{n=1}^{N} f_n S_{Vn}/V_n^2 
\end{equation} 

Here $S_{Vn}/V_n^2$ is the normalized noise spectral density at $n$ different frequencies $f_n$. The other method to quantify noise amplitude is to measure normalized noise spectral density at a fixed frequency, however this method is more prone to errors\cite{lin2008strong}.\\
Fig. 2(b) and Fig. 2(c) shows the noise amplitude as a function of effective carrier density ($n_{eff} = n_{TG}-n_{Dirac}$) beneath the top gate for different values of $n_{BG}$ with $n_{BG} >0$. At low $n_{BG}$ (close to back gate Dirac point) the noise is almost independent of effective carrier concentration. As the back gate density is increased, the noise amplitude starts increasing. The noise amplitude shows a clear "M" shape with effective density at $n_{BG} = 1.3 \times 10^{12} cm^{-2}$. With the further increase in $n_{BG}$, noise amplitude increases, and obtains a maximum peak value of $ \approx 1.1 \times 10^{-7}$ at $n_{BG} = 1.5 \times 10^{12} cm^{-2}$. Further increase in $n_{BG}$ leads to an over all decrease in noise amplitude. Our measurements show similar trend for negative values of $n_{BG}$ (SM Fig. S4).  To elucidate the noise behaviour with back gate carrier concentration, we plot noise amplitude with $n_{BG}$ in Fig. 3. It is obtained by finding maximum noise amplitude value for different $n_{BG}$ from Fig. 2(b), 2(c) and SM Fig. S4.\\

The noise in graphene show various shapes and noise magnitudes\cite{kaverzin2012impurities,balandin2013low,kumar2016tunability, rehman2022nature,stolyarov2015suppression,kayyalha2015observation,kakkar2020optimal,
kamada2021suppression,aamir2019electrical,tian2020tunable,liu2013origin,
mavredakis2018understanding,peng2017carrier,kumar2015ultra,pal2010large}, they have been explained considering the charge number fluctuation or carrier mobility fluctuation model or the combination of both. According to McWhorter charge number fluctuation model the $1/f$ noise originates from charge trapping and detrapping events occurring between the channel and the trap states in gate oxides, which are located at different distances from the channel. The noise spectral density is written as\cite{dmitriev2009hooge}
\begin{equation}
 S_I/I^2 = \frac{\lambda kT N_t}{fA Vn^2} 
\end{equation}
Here $\lambda$ is the tunnelling constant, $N_t$ is the trap concentration near Fermi energy, $A$ is the gate area and $n$ is the carrier concentration in the device. 
\begin{center}
	\begin{figure}[t]
		\includegraphics[width=.5\textwidth]{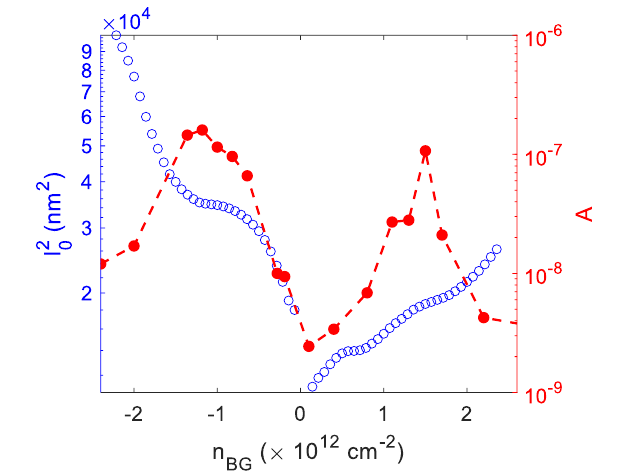}
		\caption{(color online) Noise amplitude (A) and $l_{0}^2$ on right and left y axis, respectively as a function of back gate carrier density at fixed $n_{eff} \approx 1\times 10^{11} cm^{-2}$. The noise amplitude is obtained by finding the maximum noise amplitude value for different $n_{BG}$ from Fig. 2 (b), 2(c) and SM Fig. S4. The mean free path is obtained  from $\sigma$ Vs $n_{BG}$ plot (SM Fig. S2), obtained by taking vertical cut line at $n_{TG} = 0$ from Fig. 1 (b). The noise is found to increase with increasing mean free path till a critical point beyond which noise decreases. }
		\label{fig:images}
	\end{figure}
\end{center}
It is important to note that our device has two different regions with different carrier concentration - One underneath the top gate, where the density is controlled by the combination of back gate and  top gate and other where density is determined only by the back gate. Although our measurements are performed as a function of top gate carrier density, but since the voltage probes are present in the back gate region, the noise in our device can originate from a combination of back gate and top gate regions. Since, $n_{eff}$ is kept constant in top gate region the McWhorter charge number fluctuation models predicts decrease in noise with increasing $n_{BG}$, which is in stark contrast with our experimental results which show non-monotonous density dependence (Fig. 3).\\ 
\begin{center}
	\begin{figure*}[t]
		\includegraphics[width=1\textwidth]{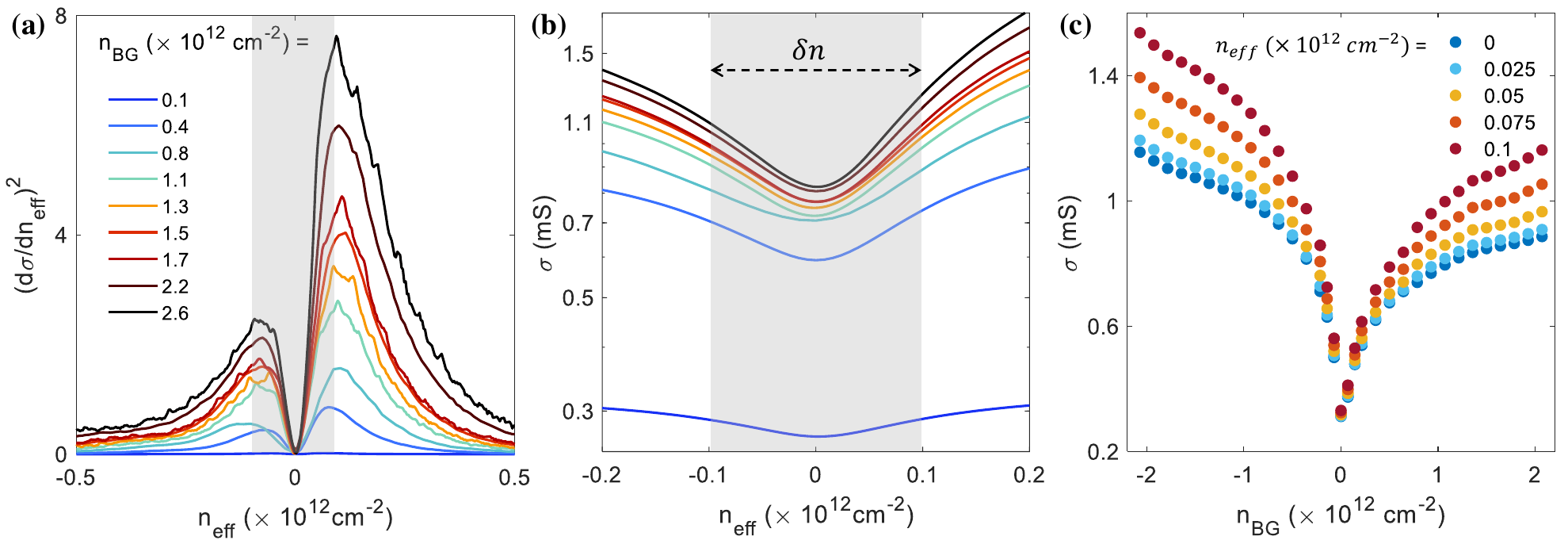}
		\caption{(color online) (a) $(d\sigma /dn_{eff})^2$ as a function of effective carrier density in the top gate region. The dip and peaks in $(d\sigma /dn_{eff})^2$ corresponds to Dirac point and charge inhomogeneity region underneath the top gate region. The estimated $\delta n \sim 1\times 10^{11} cm^{2} $ is found to be independent of $n_{BG}$. The shaded region represents the $\delta_n$ region. (b) Conductivity as a function of effective carrier density in the top gate region. The $\delta n$ region is represented by the shaded grey region. An increase in conductivity is observed even in the $\delta n$ region. (c) Conductivity as a function of $n_{BG}$ at few representative      points in $\delta n$ region, obtained by taking a vertical cut line from Fig. 4 (b). A clear increase in conductivity is observed with $n_{BG}$ even in $\delta n$ region. }
		\label{fig:images}
	\end{figure*}
\end{center}
Recent years have seen increasing evidence of mobility fluctuation in graphene devices\cite{zahid2013reduction,balandin2013low,cultrera2018role,rehman2022nature,kamada2021suppression}.
The mobility fluctuation model predicts that noise originates from the superposition of multiple events which changes the the scattering cross section $\sigma$. The power spectral density is given by\cite{balandin2013low,rehman2022nature,kamada2021suppression} 
\begin{equation}
 S_I/I^2 \propto  \frac{N_t^\mu}{V}\frac{ \tau \zeta (1-\zeta) }{ 1+ (\omega \tau)^2 } l_{0}^2 (\sigma_2 - \sigma_1)^2
\end{equation}
 where $N_t^\mu$ is the concentration of centres contributing to mobility fluctuation, $\tau$ is the characteristic time, $\zeta (1-\zeta)$ is the probability of with cross section $\sigma_1 (1-\sigma)$, $V$ is the volume of the sample and $l_{0}^2$ is the mean free path. The model predicts that noise is directly proportional to $N_t^\mu $ and  $l_{0}^2$. The mean free path ($l_{0}$) is calculated from $\sigma$ Vs $n_{BG}$ curve, obtained from Fig. 1 (a) by taking vertical cut at $n_{TG} = 0 $ (SM). The resultant $l_{0}^2$ is plotted in the left axis of Fig. 3. We find that noise amplitude, $A$ and $l_0^2$ increases with increasing $n_{BG}$.  which is consistent with the mobility fluctuation model. However, beyond a critical $l_{0}^2$, noise starts to decrease. This can be due to decrease in $N_t^\mu$ which compensate the effect of increasing $l_{0}^2$. Similar results were also reported by Cultrera et al.\cite{cultrera2018role}.\\ 
The enhanced mean free path through out the device should also increase the conductivity in the top gated region. For this 
we plot $(d\sigma/dn_{eff})^2$ for different $n_{BG}$ in Fig. 4 (a). As can be seen from the figure, the charge inhomogeneity region is almost independent of $n_{BG}$, for $n_{BG} > 0.8 \times 10^{12} cm^{-2}$. We estimate $\delta n \approx 1 \times 10^{11} cm^{-2}$, this is highlighted by grey stripe. In fig. 4 (b), we plot conductivity as a function of $n_{eff}$ for different $n_{BG}$ close to Dirac point. We find an increase in conductivity even in $\delta n$ region. 
To highlight the same, we plot conductivity in $\delta n$ region as a function of $n_{BG}$ in Fig. 4 (c). It is obtained by taking vertical line traces from Fig. 4 (b) in $\delta n$ region. A clear increase in conductivity is found in $\delta n$ region. However, by definition the conductivity in $\delta n$ region should not change. The increase in conductivity in $\delta n$ region underneath the top gate can arise from an overall increase in device mean free path (Fig. 3), which is consistent with the mobility fluctuation model.

\begin{center}
	\begin{figure*}[t]
		\includegraphics[width=1\textwidth]{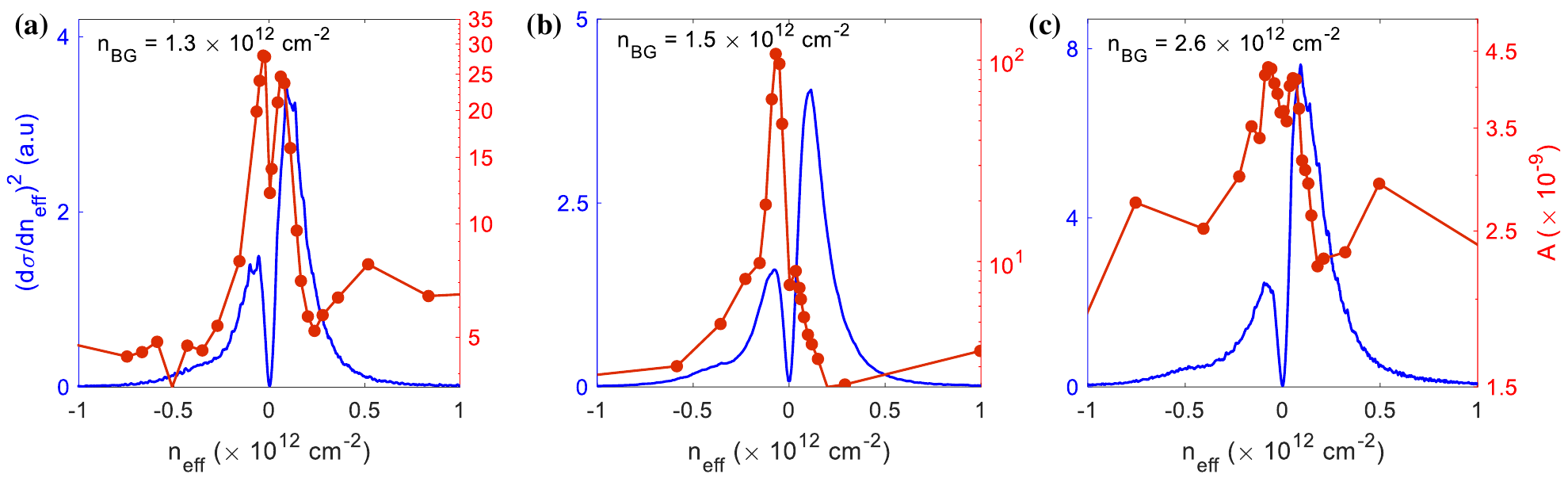}
		\caption{(color online) (a-c) Noise amplitude (A) and $(d\sigma /dn_{eff})^2$ on right and left y axis, respectively as a function of effective carrier density at few representative $n_{BG}$. The dip in $(d\sigma /dn_{eff})^2$ corresponds to Dirac point underneath the top gate region and the peaks corresponds to charge inhomogeneity region, $\delta n$. We estimate $\delta n \sim 1 \times 10^{11} cm^{-2}$. The dip observed in noise corresponds to the Dirac point while the peaks in noise is near the $\delta n$.  }
		\label{fig:images}
	\end{figure*}
\end{center}

To understand the shape and position of noise in Fig. 2 (b-c), we plot noise amplitude (A) and  $(d\sigma/dn_{eff})^2$ as a function of $n_{eff}$ for different set of back gate densities in Fig. 5 on right and left axes, respectively. The dip in $(d\sigma/dn_{eff})^2$ corresponds to Dirac point and the peaks corresponds to the charge inhomogeneity ($\delta n$) of the graphene channel underneath the top gate 
We find that the noise minimum coincides with the Dirac point while the noise peak values are close to charge inhomogeneity point of the graphene channel underneath the top gate 
(Fig. 5). Similar "M" shape noise behaviour was also reported in hBN encapsulated graphene as a function of back gate carrier density\cite{kumar2021effect}. The charge inhomogeneity model of Xu et al.\cite{xu2010effect} can qualitatively explain the position of noise maximum in Fig. 5 (a) and 5 (c). In the charge inhomogeneity region the noise amplitude can be written as combination of noise originating from the electron and hole charge puddles i.e $A \sim \alpha_H/ (n_e D_e) + \alpha_H/ (n_h D_h)$; here $n$ and $D$ represents the density and puddle size, and subscript represents electron-hole. Thus, electron doping in the channel will not only increase the electron density but will also increase the electron puddle size and decrease the hole puddle size.  This leads to less noise due to electron charge carriers and large noise due to minority charge carriers leading to increase in noise in charge inhomogeneity region. With the further increase in carrier density, the minority (hole) puddle size starts shrinking and can no longer contribute to the noise. Thus, beyond $\delta n$ region, main contribution to noise comes from majority (electron) carriers and hence noise decreases following Hooges Empirical relation, $A \sim \alpha/N_e$. The disappearance of {"M"} shape noise in Fig. 5 (b) can be due to spatial variation of charge inhomogeneity or Fermi energy fluctuation\cite{pellegrini20131}. \\

In conclusion, our measurements show that various different noise shapes and noise magnitude reported till date in hBN encapsulated devices can be realized in a single graphene device by performing noise measurement in dual gated geometry. Our results show that at low back gate density, noise is small and is almost independent of effective carrier  density. As the back gate density is increased, the noise magnitude increases by almost two orders of magnitude. With the further increase in back gate density, the noise magnitude decreases. From our analysis we trace the origin of noise emanating from combination of charge inhomogeneity and mobility fluctuation mechanism.\\

\textbf{Supplementary Material}\\
See the supplementary material for optical image of the device, mean free path calculation, $1/f$ noise measurement technique, noise contribution from top gated and non-top-gated region

\begin{acknowledgments}
The authors gratefully acknowledge National Nano Fabrication Centre (NNFC) and Micro and Nano Characterization Facility (MNCF) at Center for Nano Science and Engineering (CeNSE), IISc for help and support in carrying out this work. The authors acknowledge funding support for CeNSE facilities from the Ministry of Human Resource Development (MHRD), Ministry of Electronics and Information Technology (MeitY), and Department of Science and Technology (DST). The authors  thank Anindya Das for insightful discussion. C.K thanks IISc start up grant and QuRP seed fund grant for supporting this work.\end{acknowledgments}
\vspace{0.5 cm}

\textbf{Data Availability:}
The data that support the findings of this study are available from the corresponding author upon reasonable request.\\

\textbf{References:}

\nocite{*}
\bibliography{ref}

\end{document}